\documentclass[%
twocolumn,
prc,
]{revtex4-1}

\usepackage{graphicx}
\usepackage{dcolumn}
\usepackage{bm}
\usepackage{color} 

\usepackage{cases}
\begin{document}


\title{Precision measurement of the $E2$ transition strength to the 2$^+_1$ state of $^{12}$C}

\author{A.~D'Alessio}
\email{adalessio@ikp.tu-darmstadt.de}
 \author{T.~Mongelli}
 \email{tobias.mongelli@physik.tu-darmstadt.de}
 \author{M.~Arnold}
 \author{S.~Bassauer}
 \author{J.~Birkhan}
\author{M.~Hilcker}
\author{T.~H\"uther}
\author{J.~Isaak}
\author{L.~J\"urgensen}
\author{T.~Klaus}
\author{P.~von~Neumann-Cosel}
\author{N.~Pietralla}%
\email{pietralla@ikp.tu-darmstadt.de}
\author{V.~Yu.~Ponomarev}
\author{P.~C.~Ries}
\author{R.~Roth}
 \email{robert.roth@physik.tu-darmstadt.de}
\author{M.~Singer}%
\author{G.~Steinhilber}
\author{K.~Vobig}
\author{V.~Werner}

\affiliation{%
 Institut f\"ur Kernphysik, Technische Universit\"at Darmstadt, 64289 Darmstadt, Germany}%

\date{\today}

\begin{abstract}
The form factor of the electromagnetic excitation of $^{12}$C to its 2$^+_1$ state was measured  at extremely low momentum transfers in an electron scattering experiment at the S-DALINAC. 
A combined analysis with the world form factor data results in a reduced transition strength $B(E2; 2^+_1\rightarrow 0^+_1) =7.63(19)$ e$^2$fm$^4$ with an accuracy improved to 2.5\%. 
In-Medium-No Core Shell Model results with interactions derived from chiral effective field theory are capable to reproduce the result.
A quadrupole moment $Q(2^+_1) = 5.97(30)$ e\,fm$^2$ can be extracted from the strict correlation with the $B(\text{E2})$ strength emerging in the calculations.

\end{abstract}

\maketitle


\section{\label{sec:level1}Introduction}
\textit{}

Alpha clustering dominates the structure features of many light nuclei, especially of so-called $\alpha$-like nuclei with mass numbers $A = 4n$, where $n$ is an integer \cite{freer_2018}. 
The nucleus $^{12}$C is a prime example with the first excited $0^+$ state (the Hoyle state) showing pronounced cluster features \cite{chernykh_2007}.
Accordingly, a variety of microscopically based cluster models have been developed (see Ref.~\cite{freer_2018} and references therein).
There, the $B(E2)$ transition strength to the $2^+_1$ state plays a special role because it determines the degree of $\alpha$ clustering in the ground state (g.s.) wave function and many properties of rotational and vibrational states built on it. 
A particular example are algebraic models exploiting geometrical symmetries \cite{bijker_2020}. 

On the other hand, the nucleus $^{12}$C is a crucial testing ground for ab-initio calculations in modern theoretical nuclear physics.
The No Core Shell Model (NCSM), as well as importance truncated no-core shell model (IT-NCSM) calculations and other theoretical approaches like coupled cluster methods \cite{Eppelbaum_106,Eppelbaum_109,Dreyfuss_2013,Kravaris,Forssen_2013, PiePer2005516, maris_c_2014,barrett_ab_2013,navratil_structure_2007,navratil_recent_2009,roth_ab_2007,roth_importance_2009,tichai_natural_2019, epelbaum_ab_2011,neff_hoyle_2014,yuta_3_2016, IMNCSM2017, chernykh_pair_2010,neff_hoyle_2014,yuta_3_2016,FREER20141}  
focus on describing and predicting g.s.\ properties, excitation energies and spectroscopic quantities in $p$- and $sd$-shell nuclei. 
Since the model space increases strongly with the number of nucleons, the NCSM can be used for light nuclei only. 
To overcome this limitation, the In-Medium Similarity Renormalization Group (IM-SRG)\cite{imsrg_single_ref_heiko} has been combined with the NCSM forming the In-Medium No Core Shell Model (IM-NCSM)\cite{IMNCSM2017}, which allows to improve significantly the convergence behaviour.
Observables that react sensitively to long-range correlations of the wave function, such as radii, the quadrupole moment or the $B(E2)$ strength, converge more slowly than, for example, excitation energies. 
This makes them important for setting boundary conditions for calculations.

A remarkable correlation between the $B(E2; 2^+_1 \rightarrow 0^+_1$) strength and the quadrupole moment $Q(2^+_1)$ in $^{12}$C was observed recently for a wider range of chiral effective field theory (EFT) interactions \cite{calci_sensitivities_2016}.
Experimentally, the value of the $2^+_1$ quadrupole moment of 6(3) efm$^2$ \cite{vermeer_electric_1983} was poorly known only.
Therefore, a Coulomb-excitation reorientation-effect measurement was recently carried out \cite{kumar_raju_reorientation-effect_2018}. 
Based on the then available information for the $B(E2)$ strength, the oblate g.s.\ deformation expected from the cluster models could be confirmed but the overall uncertainty was only slightly improved to about 35\%.
The reorientation of the magnetic sub-states of the 2$^+_1$ state is a second-order process and in order to extract $Q(2^+_1)$ from the experimental data it is necessary to know the first order process (i.e.\ the $B(E2)$ strength) as precise as possible to further improve the uncertainty.

Considering the impact on the above problem and the general importance as a benchmark for the structure calculations, an improved value of the B(E2; $2^+_1 \rightarrow 0^+_1$) transition strength in $^{12}$C is clearly of interest and various experimental approaches are currently being pursued including nuclear resonance fluorescence selfabsorption experiments \cite{romig_2015} and the ($e,e^\prime$) experiment presented in this paper.

\section{\label{sec:Experiment}Electron scattering experiment}

The form factor measurements of the transition to the $2^+_1$ state of the $^{12}$C nucleus were performed with the LINTOTT spectrometer \cite{lenhardt_2006} using an electron beam of 42.5 MeV from the S-DALINAC \cite{pietralla_institute_2018} impinging on a 100 mg/cm$^2$ natural carbon target (98.9\% abundance of $^{12}$C). 
The LINTOTT spectrometer was placed at angles of 69$^\circ$, 81$^\circ$ and 93$^\circ$ with respect to the incoming electron beam, allowing measurements at extremely low momentum transfers of $q \simeq (0.25-0.32)$ fm$^{-1}$.
The low-$q$  data permit an improved extrapolation of the form factor of the $2^+_1$ state to the photon point ($k=E_{\rm x}/\hbar c$) as discussed below.

\begin{figure}
\includegraphics[width=\columnwidth]{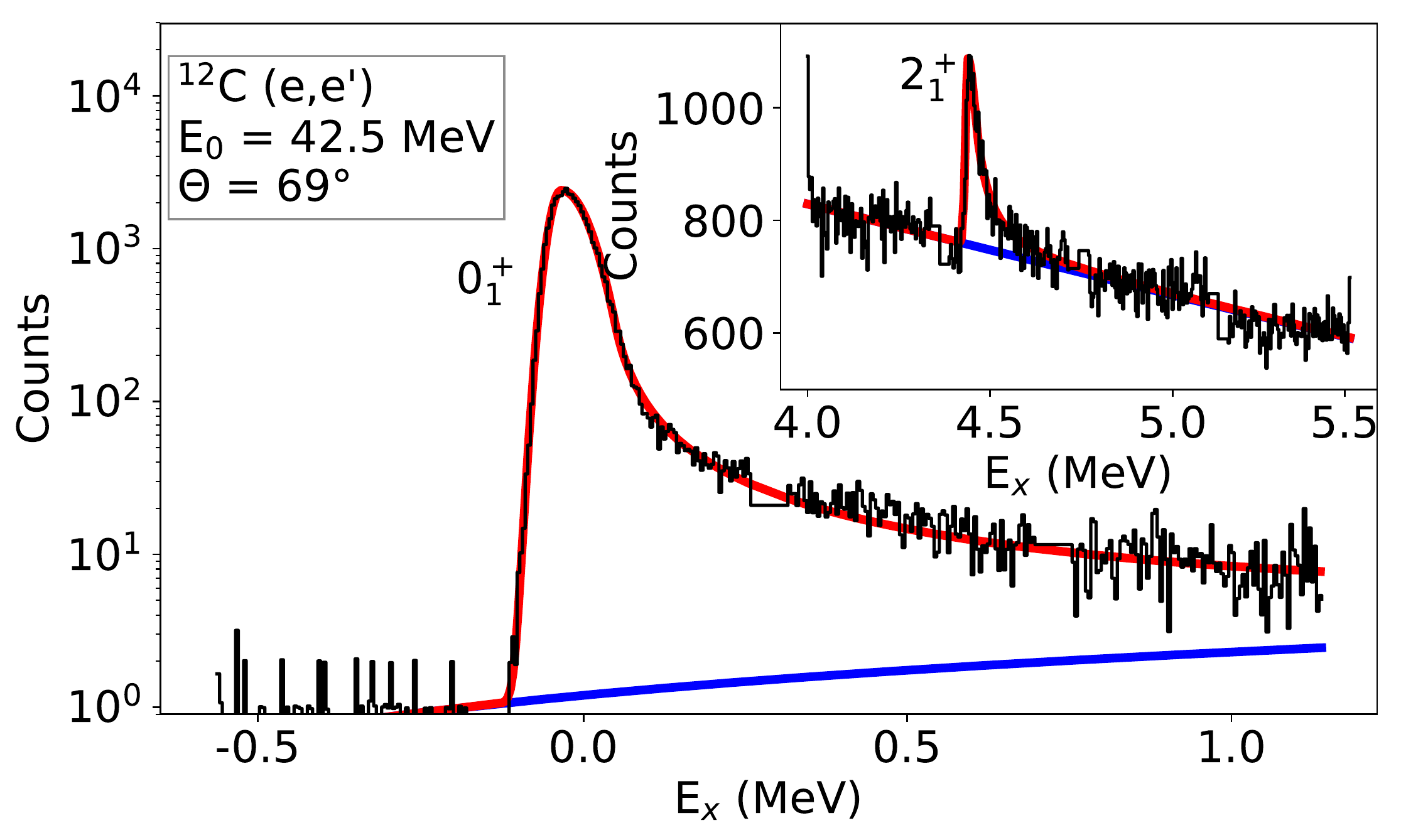} 
\caption{Elastic electron scattering spectrum taken at a beam energy of 42.5 MeV and a scattering angle of 69$^\circ$. 
The inset shows the excitation of the $2^+_1$ state.
The red lines display a fit using Eq.~(\ref{Fit}) and the blue lines a linear background.}
\label{fig:spectra}
\end{figure}

Since elastic scattering cross sections in $^{12}$C are known with high precision \cite{crannell_elastic_1966,sick_elastic_1970,reuter_nuclear_1982,kline_elastic_1973,jansen_nuclear_1972}, the form factor of the excited 2$^+_1$ state was determined in a relative measurement.
At the low beam energy, the momentum acceptance of the spectrometer of 2\% is not sufficient to observe the g.s.\ and the excited state transition with the same magnetic field settings.
However, the fields can be set in such a way that the peaks of the ground state and of the $2^+_1$ state appear in the same channels of the silicon strip focal plane detector \cite{lenhardt_2006} minimizing solid angle and efficiency uncertainties of the detector system. 
An example of the elastic scattering data is shown in Fig.~\ref{fig:spectra}.
The inset presents  a corresponding measurement of the excitation of the $2^+_1$ state.

\begin{figure}[ht]
\includegraphics[width=\columnwidth]{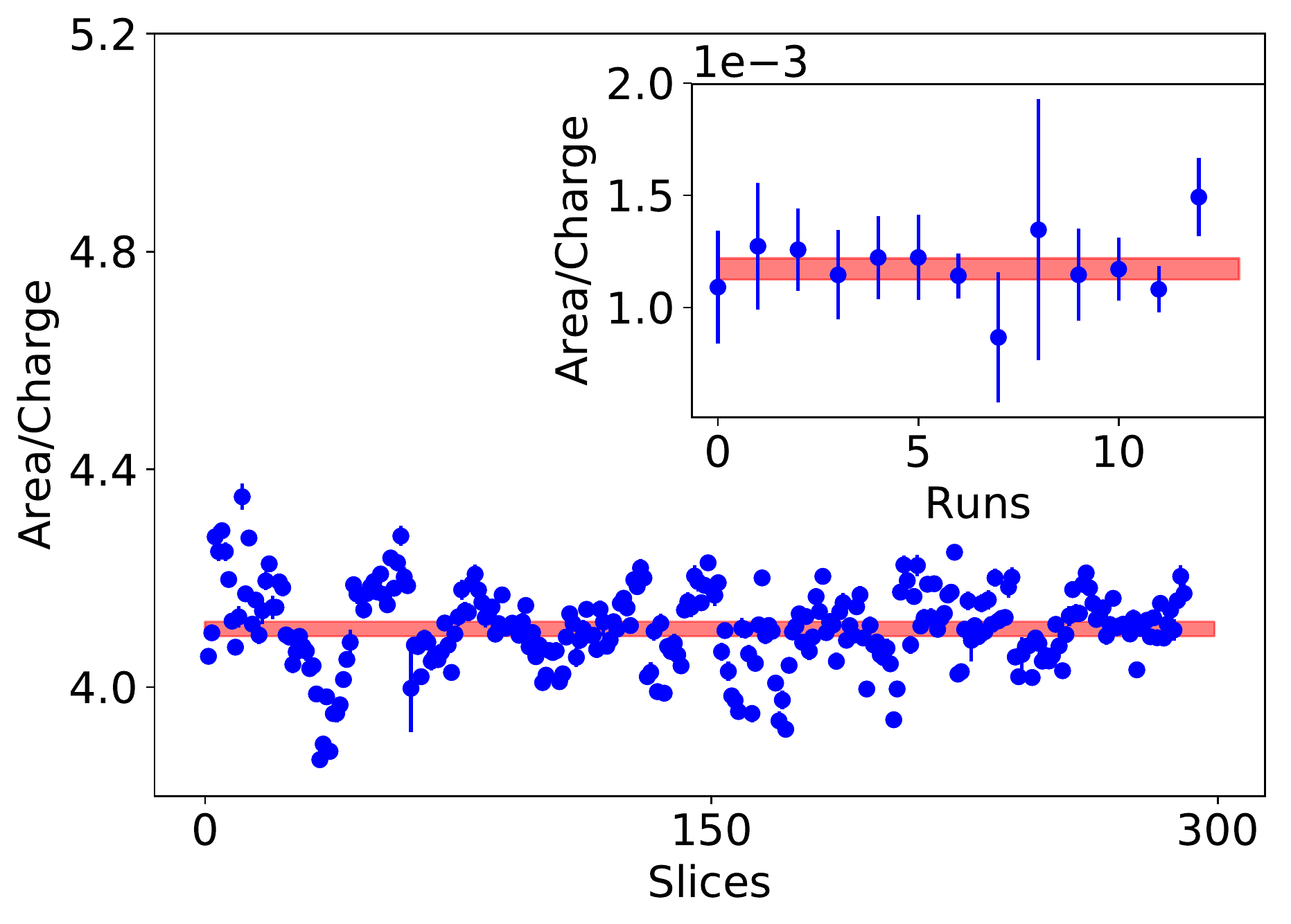} 
\caption{Area/Charge ratios for slices of the elastic line measurements at $69^\circ$ (blue circles) and uncertainty weighted average (red band).
The inset shows corresponding values for the excitation of the 2$^+_1$ state.}
\label{fig:weighted}
\end{figure}

In order to further reduce the systematic uncertainties, the data taking for the inelastic transition was stopped in regular intervals and intermittent measurements of the elastic line were performed. 
Thus, variations due to possible changes in beam position and/or beam energy were reduced by averaging over the ratio of the peak areas normalized to the collected charge.
The elastic scattering data were sliced into spectra with 50000 counts in total before the area-over-charge ratio was determined. 
Typical fluctuations (blue circles) and the uncertainty- weighted average (red bands) for the $69^\circ$ data as an example are presented in Fig.~\ref{fig:weighted} for elastic (main figure) and inelastic (inset) scattering. 
The weighted average values are $4.107(1)$ Counts/nC for elastic scattering and 1.17(5)$\cdot10^{-3}$ Counts/nC for the inelastic scattering data.
 
The peak areas were determined by a fit using the phenomenological parameterization \cite{hofmann_2002}
\begin{equation}\label{Fit}
\small{
y(x)=y_0}\left\{\small{
                \begin{array}{ll}
                  \exp\left[-\ln 2\cdot(x-x_0)^2/\Delta x_1^2\right]  x<x_0\\
                  \exp\left[-\ln 2\cdot(x-x_0)^2/\Delta x_2^2\right]  x_0 < x \leq x_0+\eta\Delta x_2\\
                  A/(B+x-x_0)^\gamma\qquad \qquad\, \,\,\,\, x>x_0+\eta\Delta x_2
                \end{array}}
              \right.            
\end{equation}
with $x_0$ denoting the peak energy, $y_0$ the count rate at $x_0$, and $\Delta x_{1,2}$ the half widths at half maximum for $E_x < x_0$ and $E_x > x_0$, respectively. 
The parameters $\eta$, A, B, and $\gamma$ describe the radiative tail. 
A possible instrumental background was allowed for, approximated by a linear function.
The peak area was determined by integration of the deduced line shape from $x_0 - 2\Delta x_1$ to  $x_0 +5\Delta x_2$. 
Then, the form factor of the inelastic transition to the $2^+_1$ state can be determined from the relation 
\begin{equation}
|F(q)|^2_{2^+_1} = |F(q)|^2_{\text{g.s.}} \frac{A_{2^+_1}}{A_{\text{g.s.}}},
\end{equation}
where $A_{\text{g.s.}}$ and $A_{2^+_1}$ denote the areas under the peaks normalized to the collected charge of the respective measurement.
The results are summarized in Tab.~\ref{Formfactors}.
\begin{table}[b]
\caption{Experimental form factors for the transition to the $2^+_1$ state of $^{12}$C from the present experiment.}
\label{Formfactors}
\begin{tabular}{cccccc}
\toprule
$E_0$  && $\Theta_{\rm lab}$ && $q$  & $|F(q)|^2$ \\
(MeV) && (deg) &&  (fm$^{-1}$) & ($10^{-4}$) \\ 
\hline
42.5 && 93$^\circ$ && 0.322 &6.34(9)  \\ 
42.5 && 81$^\circ$ && 0.290 &4.18(7)  \\ 
42.5 && 69$^\circ$ && 0.252 &2.50(11)  \\ 
\botrule
\end{tabular}
\end{table}

Extensive form factor data have been measured for this transition over a wide range of momentum transfers, but not below $q = 0.405$ fm$^{-1}$ \cite{fregeau_elastic_1956, crannell_determination_1964, fregeau_high-energy_1955, bernheim_study_1967}.
In Ref.~\cite{chernykh_pair_2010} an analytic, global, and model-independent analysis of transition form factors of exited states was introduced. 
\begin{equation}
F(q) = \frac{1}{Z}e^{-\frac{1}{2}(bq)^2}\sum_{n=1}^{n_{\text{max}}} c_n(bq)^{2n}, 
\label{eq:ff-th}
\end{equation}
with $Z$ being the charge of the probed nucleus, $q$ the momentum transfer of the electron, and $b$, $c_n$ fit parameters.
As illustrated in Ref.~\cite{chernykh_pair_2010} for the example of the transition to the $0^+_2$ state (the Hoyle state), inclusion of low-$q$ data is essential for a minimization of uncertainties.

\begin{figure}
\includegraphics[width=\columnwidth]{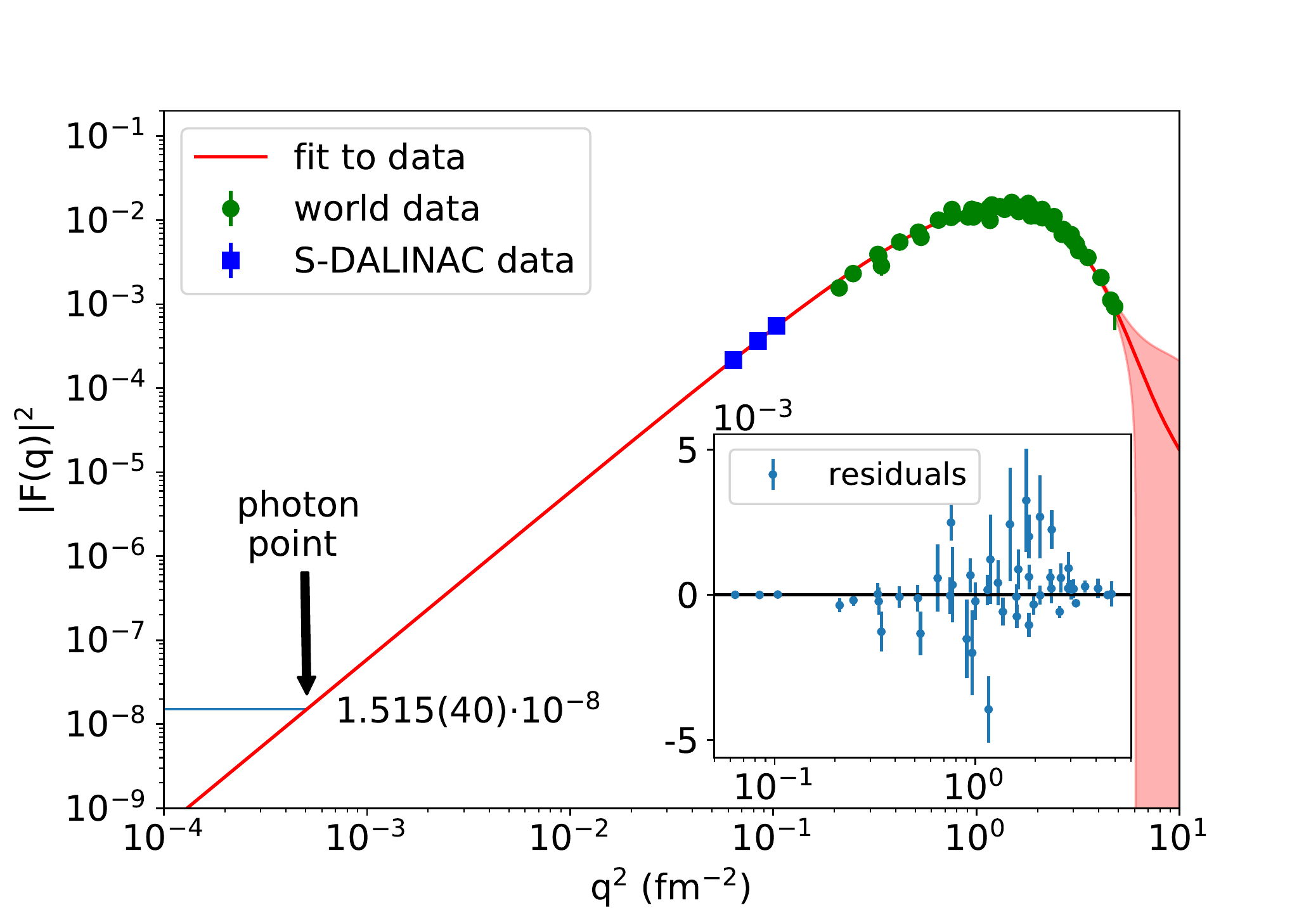} 
\caption{
Experimental form factor of the transition to the $2^+_1$ state in $^{12}$C after the DWBA corrections described in the text.
Data from the present work are shown as blue squares and previous measurements \cite{ fregeau_elastic_1956, crannell_determination_1964, fregeau_high-energy_1955, bernheim_study_1967} as green circles.
Most of the error bars are smaller than the displayed data points.
The red band shows a fit of Eq.~(\ref{eq:ff-th}) with a 1$\sigma$ uncertainty. 
The arrow indicates the photon point.
\label{fig:PhotonPointFit}
}
\end{figure}

Since Eq.~(\ref{eq:ff-th}) holds in plane wave Born approximation only, the experimental data corresponding to distorted wave Born approximation (DWBA) form factors must be corrected as outlined in Ref.~\cite{chernykh_pair_2010}.
The theoretical transition density of the $2^+_1$ state needed as starting point of the iterative procedure stems from a NCSM calculation. 
Figure \ref{fig:PhotonPointFit} presents the corrected experimental form factor data together with a fit of Eq.~(\ref{eq:ff-th}) shown as red band. 
The results of Ref.~\cite{crannell_elastic_1966} at very high $q$ with incident energies of $600 - 800$ MeV were not taken into account as it was not possible to calculate a DWBA correction for these data and their contribution to the extrapolation of the transition form factor to the photon point is negligible.

The fit provides a value $|F(q)|^2 = 1.515(40) \cdot 10^{-8}$ at the photon point. 
The impact of the current experiment can be seen from the corresponding result obtained without the low-$q$ data points $|F(q)|^2 = 1.443(161) \cdot 10^{-8}$ with a four times larger relative uncertainty.
Using the relation \cite{theissen_spectroscopy_1972} 
\begin{equation}
B(E2; 2^+_1 \rightarrow 0^+_1) = \frac{45 Z^2}{4 \pi q^4} \lim_{q\rightarrow k} |F(q)|^2
\end{equation}
we derive a transition strength of 7.63(19) e$^2$fm$^4$.
This agrees with the literature value 7.94(40) e$^2$fm$^4$ \cite{pritychenko_tables_2016} within error bars but improves the uncertainty from currently 5.5\% to 2.5\%.

\section{\label{sec:Theory}In-Medium NCSM Calculations}

For the theoretical description of the spectroscopy of $^{12}$C we use the IM-NCSM introduced in Ref.~\cite{IMNCSM2017}. 
This novel \emph{ab initio} method combines NCSM \cite{NCSM_2000, NCSM_2009} with an IM-SRG \cite{Hergert_2013, Hergert_2016, Hergert_2018} decoupling of the many-body Hamiltonian, which drastically accelerates the model-space convergence of the NCSM. 
This is particularly relevant for the description of electric quadrupole observables for nuclei in the upper $p$-shell and above, as these observables cannot be fully converged within the standard NCSM or the IT-NCSM \cite{Forssen_2013,MaVa14,calci_sensitivities_2016}. 

The IM-NCSM calculation is a four-step process: In a first step, an optimized single-particle basis is constructed for the nucleus and interaction under consideration, using natural orbitals for a perturbatively improved one-body density matrix \cite{Natorb}. 
In the second step, the reference state for the IM-SRG decoupling is obtained from a NCSM calculation in a small $N_{\text{max}}^{\text{ref}}$ model space. 
The third step then uses a multi-reference version of the IM-SRG using the White generator \cite{GeneratorWhite} to decouple the reference space from all excitations. 
We employ the Magnus formulation of the flow equations, which enables a consistent and efficient transformation of the Hamiltonian and all other operators, including the electric quadrupole operator \cite{PhDVobig}. 
In the final step, the IM-SRG-transformed operators are used in a NCSM calculation for moderate $N_{\max}$. 
The two model-space truncation parameters, $N_{\text{max}}^{\text{ref}}$ and $N_{\text{max}}$, will be used later on for the quantification of uncertainties in this many-body approach.

All calculations build on a new family of chiral two- plus three-nucleon interactions presented in Ref.~\cite{EMN-Int}.
Starting from the accurate chiral two-nucleon interactions by Entem, Machleidt, and Nosyk \cite{EMN} with non-local regulators up to N$^3$LO for three different cutoffs $\Lambda = 450\,\text{MeV/c}$, $500\,\text{MeV/c}$, and $550\,\text{MeV/c}$, we supplement chiral three-body forces at N$^2$LO and N$^3$LO with the same regulators and cutoff values. 
The low-energy constants in the three-nucleon sector are determined from the $^3$H and the $^{16}$O ground state energies. 
This leads to a family of interactions that provides a good simultaneous description of ground state energies and charge radii up into the medium-mass regime and, at the same time, a good description of excitation spectra of light nuclei \cite{EMN-Int}. 
The Hamiltonian is evolved in a free-space SRG evolution at the three-body level with a flow-parameter $\alpha$ = 0.04 fm\textsuperscript{4} \cite{RoLa11,RoCa14}. 
We note that for the $E2$ operator, we have not yet included the consistent two-body current contributions from chiral effective field theory as well as the consistent free-space SRG evolutions. 
Both are expected to have small effects on the $B(E2)$ value, smaller than our present theory uncertainties, but are eventually needed for a fully consistent description.  

\begin{figure}[t]
\includegraphics[width=\columnwidth]{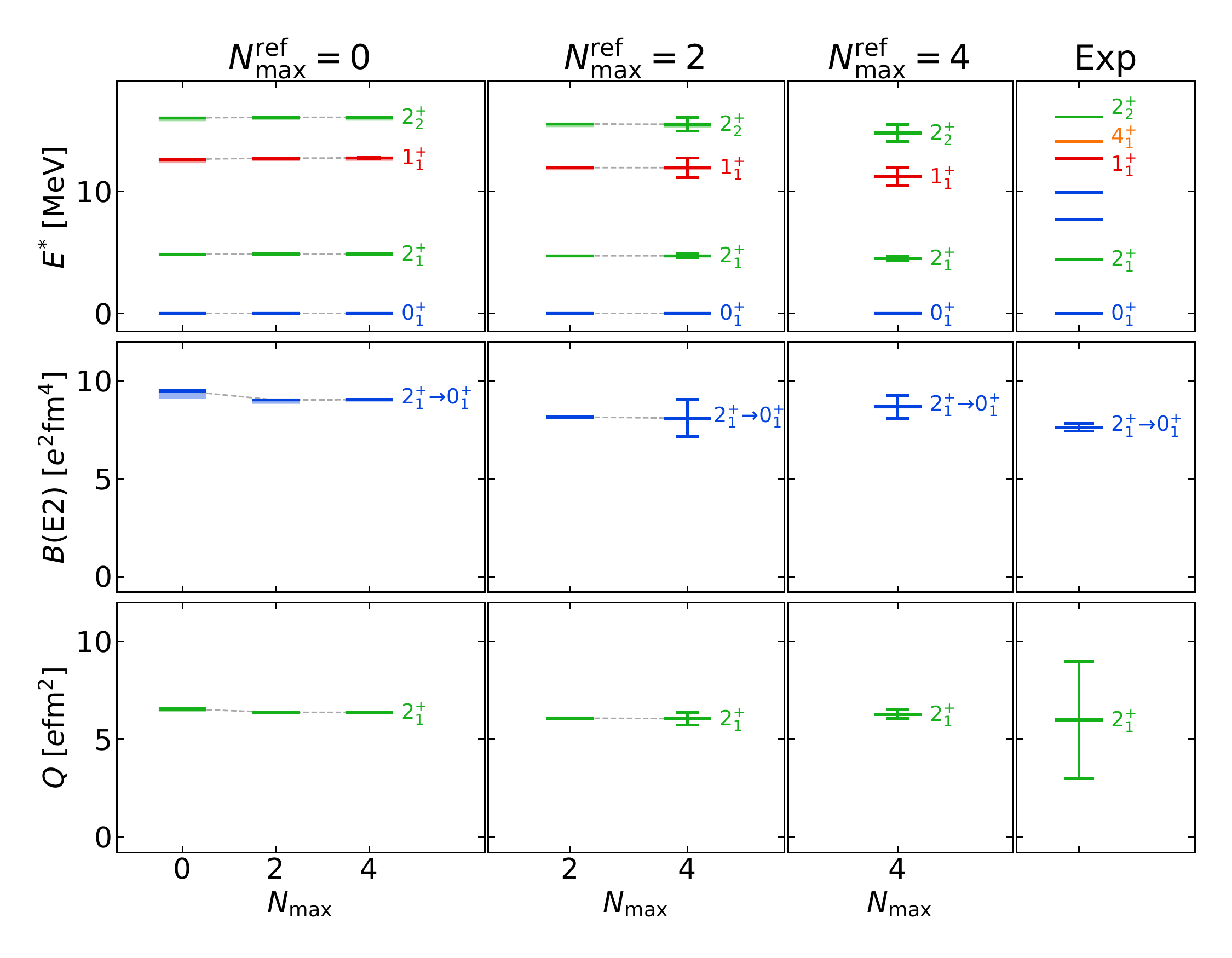} 
\caption{Excitation spectrum, $B(E2)$ transition strength, and quadrupole moment for $^{12}$C obtained in the IM-NCSM for different reference-space truncations $N_\mathrm{max}^\mathrm{ref}$ (panels left to right) as function of $N_{\text{max}}$. 
All calculations are performed with the chiral two- plus three-body interaction at N$^3$LO with cutoff $\Lambda=500$ MeV/c. 
The error bars indicate the many-body uncertainties (see text).
}
\label{fig:ObsNmax500N3LO}
\end{figure}

\begin{figure}
\includegraphics[width=\columnwidth]{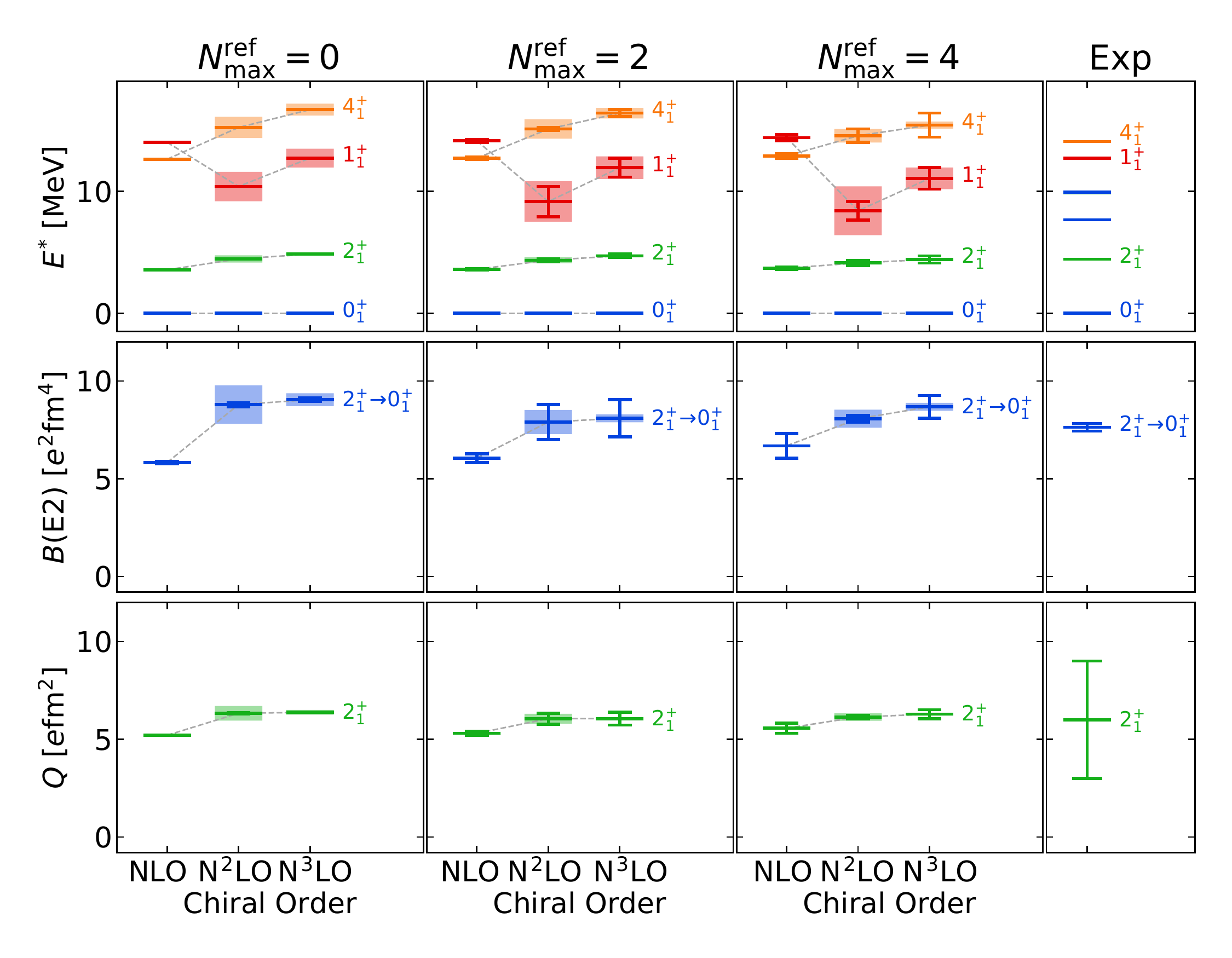} 
\caption{Excitation spectrum, $B(E2)$ transition strength, and quadrupole moment for $^{12}$C obtained in the IM-NCSM for different reference-space truncations $N_\mathrm{max}^\mathrm{ref}$ (panels left to right) with interactions from NLO to N$^3$LO with cutoff $\Lambda=500$ MeV/c. 
The error bars represent many-body uncertainties and the shaded bars indicate the interaction uncertainties (see text).}
\label{fig:C12_Obs_Int}
\end{figure}

To illustrate the superior convergence behavior and the uncertainties of the IM-NCSM calculation, Fig. \ref{fig:ObsNmax500N3LO} depicts the excitation spectrum, the $B(E2,2_1^+\rightarrow0^+_1)$ strength, and the electric quadrupole moment $Q(2_1^+)$ as a function of $N_{\text{max}}$ for different values of $N^{\text{ref}}_{\text{max}}$. 
Obviously, the results for all observables are very stable with increasing $N_{\text{max}}$, showing that the final NCSM calculation is fully converged even for these small model spaces. 
The dependence on the reference-space size $N^{\text{ref}}_{\text{max}}$, which indirectly probes the effect of omitted normal-ordered three-body terms in the IM-SRG, is also quite small. 
We estimate the uncertainties of the many-body treatment based on the differences of the observables for successive values of $N_{\text{max}}$ and $N^{\text{ref}}_{\text{max}}$ and we also include a variation of the IM-SRG flow parameter by a factor of two. 
The maximum of these three differences gives the many-body uncertainty inducted by the error bars in Fig.~\ref{fig:ObsNmax500N3LO}. 
Note that in all cases, the change of $N^{\text{ref}}_{\text{max}}$ determines this maximum and, thus, the total many-body uncertainty. 
For the interaction employed in Fig.~\ref{fig:ObsNmax500N3LO}, the chiral interaction at N$^3$LO with $\Lambda = 500\,\text{MeV/c}$ the agreement of the $2_1^+$ excitation energies, the $B(E2)$ strength, and the quadrupole moment with experiment is remarkable. Moreover, the new family of chiral interactions gives us the opportunity to study the robustness of the results under variation of the chiral order. 
This is illustrated in Fig.~\ref{fig:C12_Obs_Int} for the interactions from NLO to N$^3$LO with cutoff $\Lambda = 500\,\text{MeV/c}$. 
Given the complete convergence with $N_{\text{max}}$ we only show the results for $N_{\text{max}}=4$ with error bars indicating the many-body uncertainties as described before. From the order-by-order behavior of the individual observables we can extract the uncertainties caused by the truncation of the chiral expansion. 
We use a simple prescription described in Ref.~\cite{EMN-Int}, which goes back to Refs.~\cite{EpKr15,BiCa16,BiCa18}, using the differences of subsequent orders weighted by powers of the expansion parameter. 
These interaction uncertainties at N$^2$LO and N$^3$LO are indicated by shaded bands in Fig.~\ref{fig:C12_Obs_Int}. We observe that the results for the $2_1^+$ excitation energy and the $B(E2,2_1^+\rightarrow0^+_1)$ strength robustly agree with experiment within uncertainties at N$^2$LO and N$^3$LO. 
Furthermore, we obtain an accurate prediction for the quadrupole moment with theory uncertainties that are almost an order of magnitude smaller than the present experimental uncertainties \cite{kumar_raju_reorientation-effect_2018}. 
%

\begin{figure}[t]
\includegraphics[width=0.8\columnwidth]{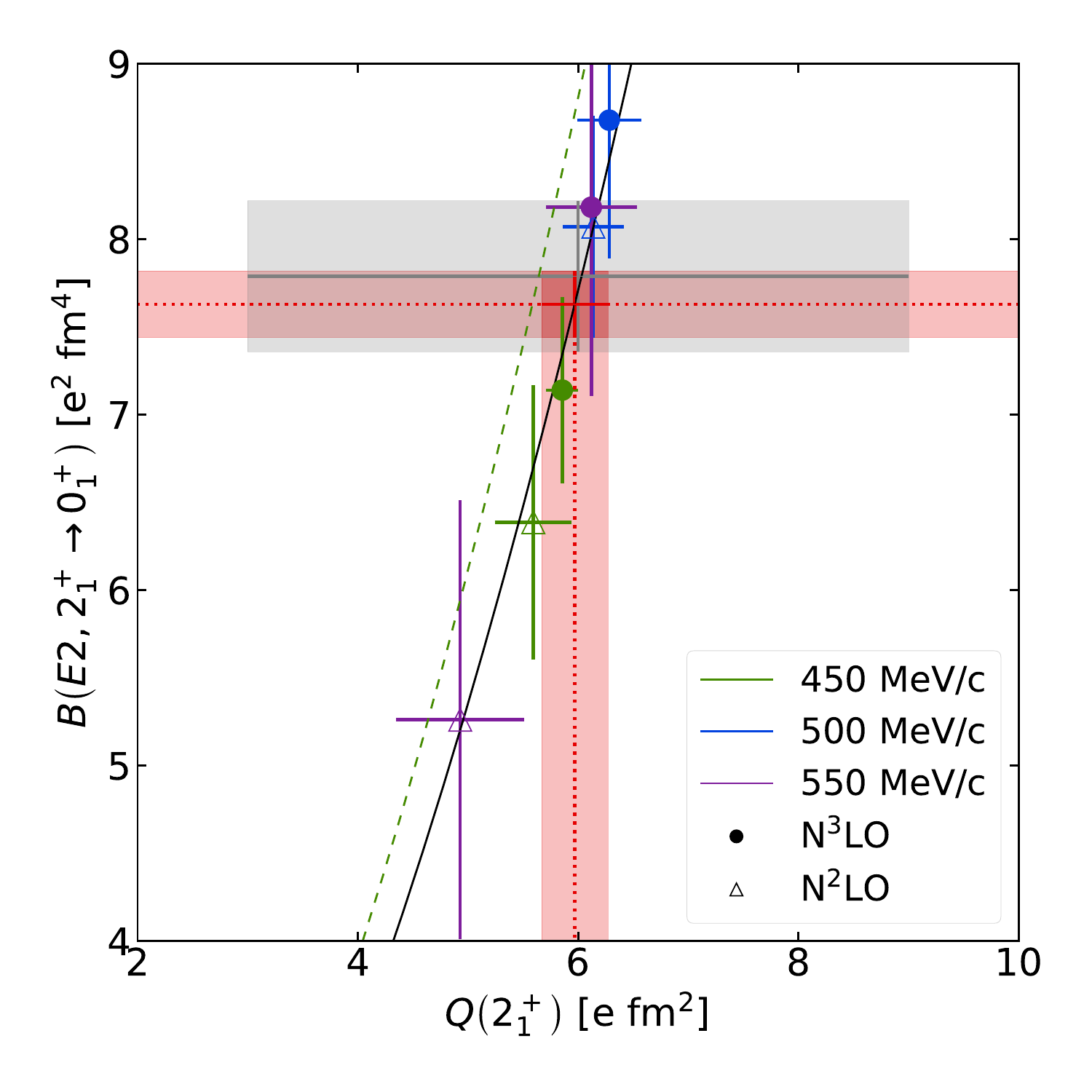}
\caption{Correlation of the quadrupole observables  $B(E2,2_1^+\rightarrow0^+_1)$ and $Q(2^+_1)$ for $^{12}$C obtained with N$^2$LO and N$^3$LO interactions for three different cutoffs. 
All IM-NCSM calculations are performed with $N_{\text{max}}=N_{\text{max}}^{\text{ref}}=4$. 
The error bars indicate the combined many-body and interaction uncertainties.
The lines show the prediction of a simple rigid rotor (dashed) and a fitted (solid) rotor model, see text.
The horizontal and vertical red shaded bands indicate the experimental $B(E2)$ value and the $Q(2^+_1$) value derived from the intersection with the model correlation. 
The grey and red areas indicate the experimental limits from literature values \cite{pritychenko_tables_2016,kumar_raju_reorientation-effect_2018} and from the present work, respectively.
} 
\label{fig:BE2Q}
\end{figure}

\begin{table}[b]
\caption{Electric quadrupole obervables obtained with the IM-NCSM for $N_{\text{max}}=N^{\text{ref}}_{\text{max}}=4$ using the N$^3$LO interactions with three different cutoffs $\Lambda$. The uncertainties include many-body and interaction uncertainties.}
\label{tab:theory}
\begin{tabular}{ccccccc}
\toprule
$\Lambda$ && $E_{\rm x}(2^+_1)$  && $B(E2,2_1^+\rightarrow0^+_1)$ && $Q(2^+_1)$ \\ 
(MeV/c) && (MeV) && ($e^2\text{fm}^4$)  && ($e\,\text{fm}^2$)  \\ 
\hline
450 && 3.96(20) && 7.14(53) && 5.86(15)  \\ 
500 && 4.41(30) && 8.68(79) && 6.28(29)   \\ 
550 && 4.45(27) && 8.18(108) && 6.12(41)  \\ 
\botrule
\end{tabular}
\end{table}
Finally, we combine the results for $B(E2,2_1^+\rightarrow0^+_1)$ and $Q(2_1^+)$ in a correlation plot shown in Fig.~\ref{fig:BE2Q}. 
We include the N$^2$LO and N$^3$LO interactions for all three values of the cutoff with error bars reflecting the combined many-body and interaction uncertainties. 
Here we only show the IM-NCSM calculations for the largest model space with $N_{\text{max}}=N^{\text{ref}}_{\text{max}}=4$. 
The results for all 6 interactions fall onto a single line, as was already observed in Ref.~\cite{calci_sensitivities_2016} for various first-generation chiral interactions. 
While N$^2$LO interactions show a larger cutoff dependence, the N$^3$LO results bracket the experimental $B(E2)$ value and show a reduced cutoff dependence, as summarized in Tab.~\ref{tab:theory}. 
The various microscopic results can be fit by a simple rotor-model correlation. 
The two lines show the correlation predicted by a rigid rotor (dashed) and the fitted rotor model with a ratio of the intrinsic quadrupole moments $Q_{0,t}/Q_{0,s}$ = 0.967 (solid). 
Details can be found in Ref.~\cite{calci_sensitivities_2016}, where almost the same ratio of the transition and static intrinsic quadrupole moments was found based on a completely different set of interactions.

%
We can combine this correlation with the new experimental value for the $B(E2,2_1^+\rightarrow0^+_1)$ to obtain an accurate value for the quadrupole moment $Q(2_1^+) = 5.97(30)\, e\,\text{fm}^2$, where the uncertainties include the average many-body and interaction uncertainties of the N$^3$LO calculations for the quadrupole moment and the experimental uncertainties of the transition strength propagated via the correlation. 
This value is compatible within uncertainties with the $Q(2_1^+)$ computed directly in the IM-NCSM with the  N$^3$LO interactions for all three cutoffs, as seen in Tab.~\ref{tab:theory}. 
The red area in Fig.~\ref{fig:BE2Q} indicates the new experimental value of the $B(E2)$ and the quadrupole moment of the $2^+_1$ state in $^{12}$C extracted from the correlation analysis, both with their uncertainties, in comparison to the literature values \cite{pritychenko_tables_2016,kumar_raju_reorientation-effect_2018} (grey area).

\section{Summary}

The present work reports a new measurement of the electron scattering form factor of the transition to the 2$^+_1$ state in $^{12}$C at very low momentum transfers.
Combined with the world data this permits an extraction of the $B(E2)$ strength based on the model-independent analysis introduced in Ref.~\cite{chernykh_pair_2010} with a much improved relative uncertainty of 2.5\%.
This highly precise value is used to benchmark a new family of chiral two- plus three-nucleon interactions \cite{EMN-Int} and test the convergence properties of calculations with the novel {\it ab initio} IM-NCSM method \cite{IMNCSM2017}.
Very good agreement is obtained.
The correlation between the $B(E2)$ and $Q(2^+_1)$ values in the model results, which can be described by a simple rotor model, permits an extraction of the hard-to-measure quadrupole moment \cite{kumar_raju_reorientation-effect_2018}  with a precision improved by almost an order of magnitude. 

\begin{acknowledgments}
This work was funded by the Deutsche Forschungsgemeinschaft
(DFG, German Research Foundation) under grant No.\ SFB 1245 (project ID 279384907) and GRK 2128 (project ID 264883531). 
The \textit{ab initio} calculations were performed on the LICHTENBERG 
high performance cluster at the computing center of the TU Darmstadt.

\end{acknowledgments}

\bibliographystyle{aapmrev4-2}
\bibliography{Biliography}
\end{document}